\documentclass[twocolumn]{autart}    

\makeatletter
\def\@biblabel#1{}
\makeatother

\usepackage{hyperref} \hypersetup{colorlinks=true, linkcolor=blue, breaklinks=true, urlcolor=blue}
\usepackage{amsmath}
\usepackage{graphicx,mathrsfs,amssymb,cite,enumerate,mathrsfs,balance,float}
\usepackage{bbm}
\usepackage{dsfont}
\usepackage{color}
\usepackage[round]{natbib}
\usepackage{doi}

\newcommand{\until}[1]{\{1,\dots, #1\}}
\newcommand{\subscr}[2]{#1_{\textup{#2}}}

\newcommand{\setdef}[2]{\{#1 \; | \; #2\}}

\newcommand{\integernonnegative}{\ensuremath{\mathbb{Z}}_{\ge 0}}
\newcommand{\real}{\ensuremath{\mathbb{R}}}

\newcommand\oprocendsymbol{\hbox{$\square$}}
\newcommand\oprocend{\relax\ifmmode\else\unskip\hfill\fi\oprocendsymbol}

\renewcommand{\theenumi}{(\roman{enumi})}

\DeclareSymbolFont{bbold}{U}{bbold}{m}{n}
\DeclareSymbolFontAlphabet{\mathbbold}{bbold}

\newcommand{\ds}{\displaystyle}

\newcommand{\Prob}{\mathbb{P}}
\newcommand{\E}{\mathbb{E}}

\usepackage[prependcaption,colorinlistoftodos]{todonotes}

\newcommand{\mcN}{\mathcal{N}}

\begin{document}

\begin{frontmatter}
\runtitle{Convergence of Heterogeneous DW Model}  

\title{Convergence Properties of the Heterogeneous Deffuant-Weisbuch Model\thanksref{footnoteinfo}}

\thanks[footnoteinfo]{This work was supported by  the National Natural
  Science Foundation of China under grants 11688101 and 61803024,
  the National Key Basic Research Program of China (973 program) under
  grant 2016YFB0800404, and the Fundamental Research Funds for the Central Universities under grant FRF-TP-17-087A1,
 and the National Key Research and Development Program of Ministry of Science and Technology of China under
  grant 2018AAA0101002. Additionally, this
  material is based upon work supported by, or in part by, the U.~S.~Army
  Research Laboratory and the U.~S.~Army Research Office under grant
  numbers W911NF-15-1-0577.}

\author[Chen]{Ge Chen}\ead{chenge@amss.ac.cn},
\author[Su1]{Wei Su}\ead{suwei@amss.ac.cn},    
\author[Mei]{Wenjun Mei}\ead{meiwenjunbd@gmail.com},
\author[Bullo]{Francesco Bullo}\ead{bullo@ucsb.edu},
\address[Chen]{National Center for Mathematics and Interdisciplinary Sciences \& Key Laboratory of Systems and
Control, Academy of Mathematics and Systems Science, Chinese Academy of Sciences, Beijing 100190,
China}
\address[Su1]{School of Automation and Electrical Engineering, University of Science and Technology Beijing, Beijing 100083, China}  
\address[Mei]{Automatic Control Laboratory, ETH, 8092 Zurich, Switzerland}
\address[Bullo]{Department of Mechanical Engineering and the Center of Control, Dynamical-Systems and Computation, University of
California at Santa Barbara, CA 93106-5070, USA}


\begin{abstract}
  The Deffuant-Weisbuch (DW) model is a bounded-confidence opinion dynamics
  model that has attracted much recent interest. Despite its simplicity and
  appeal, the DW model has proved technically hard to analyze and its most
  basic convergence properties, easy to observe numerically, are only
  conjectures.

  This paper solves the convergence problem for the heterogeneous DW
  model with the weighting factor not less than $1/2$. We establish that, for any positive confidence bounds and initial
  values, the opinion of each agent will converge to a limit value almost
  surely, and the convergence rate is exponential in mean square. Moreover, we show that the limiting opinions of any two
  agents either are the same or have a distance larger than the confidence
  bounds of the two agents. Finally, we provide some sufficient conditions
  for the heterogeneous DW model to reach consensus.
\end{abstract}

\end{frontmatter}

\section{Introduction}

The field of opinion dynamics studies the dynamical processes regarding the
formation, diffusion, and evolution of public opinion about certain events
and object of interest in social systems.  The study of opinion dynamics
can be traced back to the two-step communication flow model in
\citep{EK-PFL:55} and the social power and averaging model in
\citep{JRPF:56}. The model by \cite{JRPF:56} was then elaborated
by~\cite{FH:59} and rediscovered by~\cite{MHDG:74}.  Other notable
developments include the model by~\cite{NEF-ECJ:90} with attachment to
initial opinions, a general influence network theory \citep{NEF:98}, social
impact theory \citep{BL:81}, and dynamic social impact theory
\citep{BL:96}.  A comprehensive review of opinion dynamics models is given
in the two tutorials~\cite{AVP-RT:17,AVP-RT:18} and the
textbook~\cite{FB:19}.

In recent years, significant attention has focused on so-called
\emph{bounded confidence} (BC) models of opinion dynamics. In these models
one individual is willing to accord influence to another only if their
pair-wise opinion difference is below a threshold (i.e., the confidence
bound).  \citep{GD-DN-FA-GW:00} propose their now well-known BC model
called the Deffuant-Weisbuch (DW) model or Deffuant model.  In this model a
pair of individuals is selected randomly at each discrete time step and
each individual updates its opinion if the other individual's opinion lies
within its confidence bound. A second well-known BC model is the
Hegselmann-Krause (HK) model \citep{RH-UK:02}, where all individuals update
their opinions synchronously by averaging the opinions of individuals
within their confidence bounds.

As reported in~\citep{JL:07b,JL:09}, simulation results for the DW model
have revealed numerous interesting phenomena such as consensus,
polarization and fragmentation.  However, the DW model is in general hard
to analyze due to the nonlinear state-dependent inter-agent topology.
Current analysis results focus on the homogeneous case in which all the
agents have the same confidence bound. The convergence of the homogeneous
DW model has been proved in \citep{JL:05a} and its convergence rate is
established in \citep{JZ-GC:15}.  Some research has considered also
modified DW models. For example, \citep{GC-FF:11} consider a generalized DW
model with an interaction kernel and investigate its scaling limits when
the number of agents grows to infinity; \citep{JZ-YH:13} generalize the DW
model by assuming that each agent can choose multiple neighbors to exchange
opinion at each time step.  Despite all this progress, the analysis of the
heterogeneous DW model is still incomplete in that its convergence
properties are yet to be established.

It is worth remarking that the analysis of the HK model is also similarly
restricted to the homogeneous case; the convergence of the heterogeneous HK
model is only conjectured in our previous work~\citep{AM-FB:11f} and has
since been established in~\citep{BC-CW:17} only for the special case that
the confidence bound of each agent is either $0$ or $1$. In general,
numerous conjectures remain open for heterogeneous bounded-confidence
models.

This paper establishes the convergence properties of the heterogeneous DW
model with the weighting factor is not less than $1/2$. We show that, for any positive confidence bounds and initial
opinions, the opinion of each agent converges almost surely to a limiting
value,  and the convergence rate is exponential in mean square. Additionally we prove that the limiting values of any two agents'
opinions are either identical or have a distance larger than the confidence
bounds of the two agents. Moreover, we show that a sufficient, and in some
cases also necessary, condition for almost sure consensus; the intuitive
condition is expressed as a function of the largest confidence bound in the
group.

The paper is organized as follows.  Section~\ref{Mod_sec} introduces the
heterogeneous DW model and our main
results. Section~\ref{pomr} contains the proofs of our results.
Finally, Section \ref{Conclusions} concludes the paper.

\section{The heterogeneous DW model and our main convergence results}\label{Mod_sec}
\renewcommand{\thesection}{\arabic{section}}

This paper considers the following  DW
model proposed in \citep{GD-DN-FA-GW:00}.  In a group of $n\geq 3$ agents, we assume each agent
$i\in\until{n}$ has a real-valued opinion $x_i(t)\in\real$ at each
discrete time $t\in\integernonnegative$.  We let
$x(t):=(x_1(t),\ldots,x_n(t))^\top$ assume, without loss of generality, that
$x(0)\in[0,1]^n$.  We let $r_{i}>0$ denote the \emph{confidence bound} of
the agent $i$ and we assume, without loss of generality, $$r_1\geq r_2\geq \cdots \geq r_n>0.$$
 We let the constant $\mu\in(0,1)$ denote the weighting factor.
We let $\mathbbm{1}_{\{\cdot\}}$ denote the indicator function, i.e., we
let $\mathbbm{1}_{\{\omega\}}=1$ if the property $\omega$ holds true and
$\mathbbm{1}_{\{\omega\}}=0$ otherwise.
 At each time
$t\in\integernonnegative$, a pair $\{i_t,j_t\}$ is independently and
uniformly selected from the set of all pairs
$\mcN=\setdef{\{i,j\}}{i,j\in\until{n},i<{j}}$. Subsequently, the opinions
of the agents $i_t$ and $j_t$ are updated according to
\begin{equation}\label{m1}
\left\{
\begin{aligned}
  x_{i_t}&(t+1) = x_{i_t}(t)\\
  &+\mu \mathbbm{1}_{\{|x_{j_t}(t)-x_{i_t}(t)|
    \leq r_{i_t}\}}(x_{j_t}(t)-x_{i_t}(t)),\\
  x_{j_t}&(t+1) = x_{j_t}(t)\\
  &+\mu \mathbbm{1}_{\{|x_{j_t}(t)-x_{i_t}(t)|
    \leq r_{j_t}\}}(x_{i_t}(t)-x_{j_t}(t)),
\end{aligned}\right.
\end{equation}
whereas the other agents' opinions remain unchanged:
\begin{gather}\label{m2}
  x_{k}(t+1)= x_{k}(t), \enspace \text{for } k\in\until{n}\setminus\{i_t,j_t\}.
\end{gather}
If $r_1=\cdots=r_n$, the DW model is called homogeneous, otherwise heterogeneous.

Previous works~\citep{JL:05a} show that the homogeneous DW model
(\ref{m1})-(\ref{m2}) always converges to a limit opinion
profile. Simulations reported in~\citep{JL:07b} show that this property
holds also for the heterogeneous case; but a proof for this statement is
lacking.  Simulations in
\citep{GD-DN-FA-GW:00,GW-GD-FA-JN:02} show that the parameter $\mu$ mainly affects
 the convergence time and so previous works \citep{JL:07b,JL:09}
simplified the model by setting $\mu=1/2$.  This paper considers the case when
$\mu\in[1/2,1)$.

Before stating our convergence results, we need to define the probability
space of the DW model.  If the initial state $x(0)$ is a deterministic
vector, we let $\Omega=\mcN^{\infty}$ be the sample space, $\mathcal{F}$ be
the Borel $\sigma$-algebra of $\Omega$, and $\Prob$ be the probability measure
on $\mathcal{F}$. Then the probability space of the DW model is written as
$(\Omega,\mathcal{F},\Prob)$.
It is worth mentioning that $\omega\in\Omega$
 refers to a particular path of agent pairs for opinion update.
 If the initial state is a random vector, we let
$\Omega=[0,1]^n \times \mcN^{\infty}$ be the sample space and, similarly to
the case of deterministic initial state, the probability space is defined by
$(\Omega,\mathcal{F},\Prob)$.

Let $\|\cdot\|$ denote the $\ell_2$-norm (Euclidean norm). The main results of this paper can be formulated as follows.

\begin{thm}{\textbf{\textup{(Convergence and convergence rate of heterogeneous DW model)}}}
  \label{Main_result}
  Consider the heterogeneous DW model (\ref{m1})-(\ref{m2})
with $\mu\in[1/2,1)$. For any initial state
    $x(0)\in[0,1]^n$,
    \begin{enumerate}
    \item there exists a random vector $x^*\in[0,1]^n$ satisfying
      $x_i^*=x_j^*$ or $|x_i^*-x_j^*|>\max\{r_i,r_j\}$ for all $i\neq j$,
      such that $x(t)$ converges to $x^*$ almost surely (a.s.) as
      $t\to\infty$, and
    \item $\ds \E
          \|x(t)-x^*\|^2\leq n c^{\lfloor \frac{t}{2(T+1)}\rfloor} +
          \frac{n}{4} \!\left(\!1\!-\frac{8\mu(1-\mu)}{n(n-1)}\right)^{\!\lfloor
            \frac{t}{2}\rfloor}$
 with $T:=(n-1)^2 (1+\lceil\log_{1-\mu} \frac{r_n}{r_1}\rceil) \lceil \frac{1-r_{n}}{(1-\mu)^2 r_{n}}\rceil$
  and $c:=1-\frac{2^{T}}{n^{T}(n-1)^{T}}$.
    \end{enumerate}
\end{thm}

\begin{figure}
  \centering
  \includegraphics[width=2.5in]{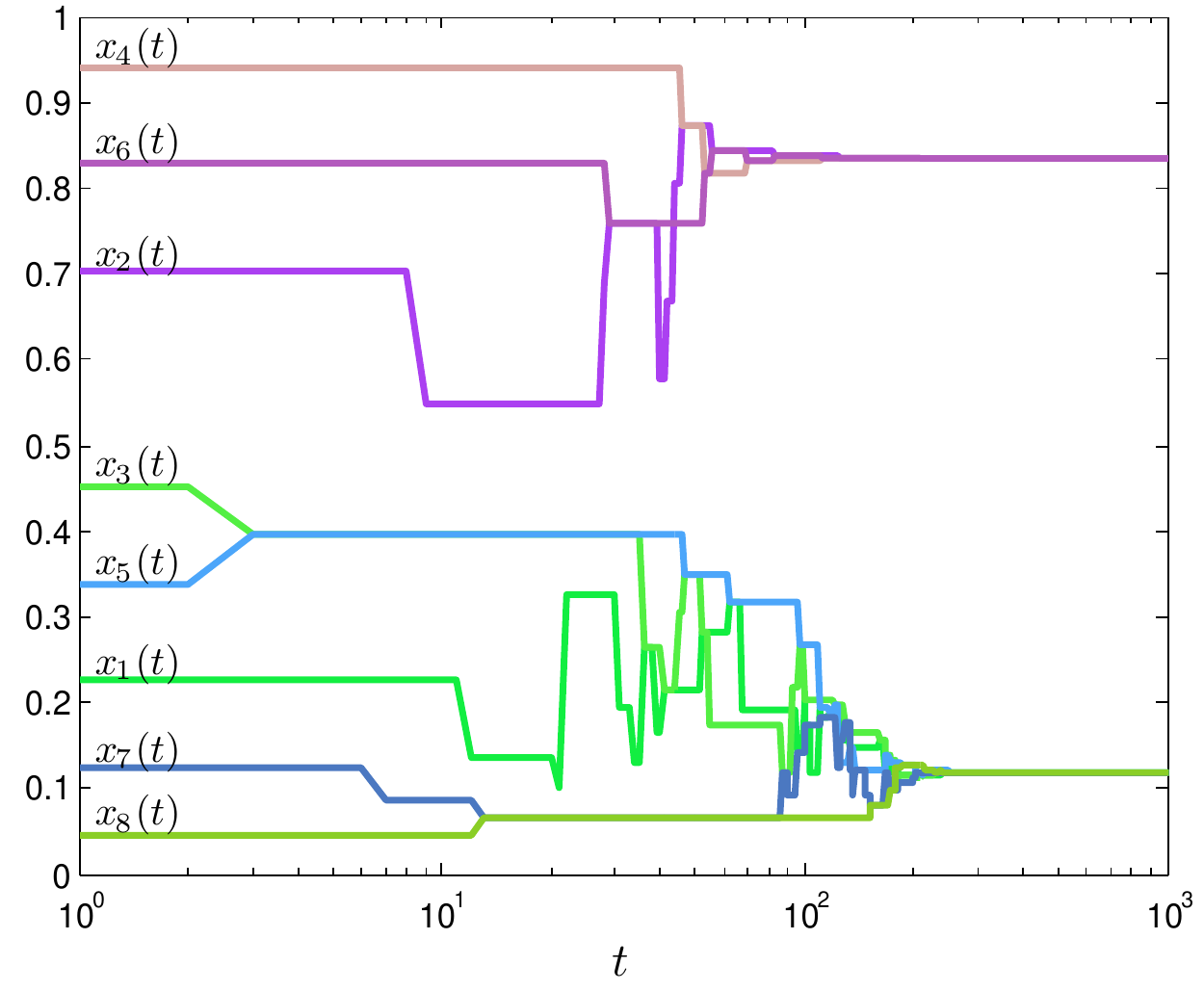}
  \caption{A convergent simulation of the heterogeneous DW
    model}\label{Cong}
\end{figure}

The proof of Theorem~\ref{Main_result} is postponed to Section~\ref{pomr}.
Fig.~\ref{Cong} displays the simulation results for a heterogeneous DW
model~\eqref{m1}-\eqref{m2} with $\mu=1/2$ and $n=8$, while the agents' confidence bounds equal to $0.5,0.41,0.35,0.24,0.175,0.165,0.12,0.047$
respectively.  Consistently with Theorem~\ref{Main_result}, Fig.~\ref{Cong}
shows that the individual opinions converge to two distinct limit values
and that the distance between the two values is larger than $r_1=0.5$.

Theorem \ref{Main_result} leads to two corollaries on convergence to
consensus. By consensus we mean that all agents' opinions converge to the
same value.

\begin{cor}{\textbf{\textup{(Almost sure consensus for large confidence bound)}}}
  \label{Cor_1}
  Consider the heterogeneous DW model (\ref{m1})-(\ref{m2}) with
$\mu\in[1/2,1)$. If the largest confidence bound $r_1$ is
not less than $1$, then for any initial state
    $x(0)\in[0,1]^n$ the system reaches consensus a.s.
\end{cor}

\newcommand{\rhomin}{\subscr{\rho}{min}}

\begin{cor}{\textbf{\textup{(Almost sure consensus if and only if large confidence bound)}}}
  \label{Cor_2}
  Consider the heterogeneous DW model (\ref{m1})-(\ref{m2}) with $\mu\in[1/2,1)$. Assume that the initial state $x(0)$ is randomly
  distributed in $[0,1]^n$ and that its joint probability density has a
  lower bound $\rhomin>0$, that is, for any real numbers
  $a_i,b_i$, $i\in\until{n}$, with $0\leq a_i<b_i\leq 1$,
  \begin{equation}\label{cor2_0}
    \Prob\Big(\bigcap_{i=1}^n \{x_i(0)\in [a_i,b_i]\} \Big)
    \geq \rhomin \prod_{i=1}^n (b_i-a_i).
  \end{equation}
  Then the heterogeneous DW model reaches consensus almost surely if and
  only if the largest confidence bound $r_1\geq 1$.
\end{cor}

\begin{rem}
  Condition (\ref{cor2_0}) can be satisfied if $\{x_i(0)\}_{i=1}^n$ are
  mutually independent and have positive probability densities over
  $[0,1]$. Examples include independent uniform, or independent truncated
  Gauss distributions. On the other hand, condition (\ref{cor2_0}) cannot
  be satisfied if there exists $x_i(0)$ which has zero probability density in
  a subinterval of $[0,1]$ with positive measure, e.g., if $x_i(0)$ is a discrete random variable.
\end{rem}

\begin{figure}
  \centering
  \includegraphics[width=2.5in]{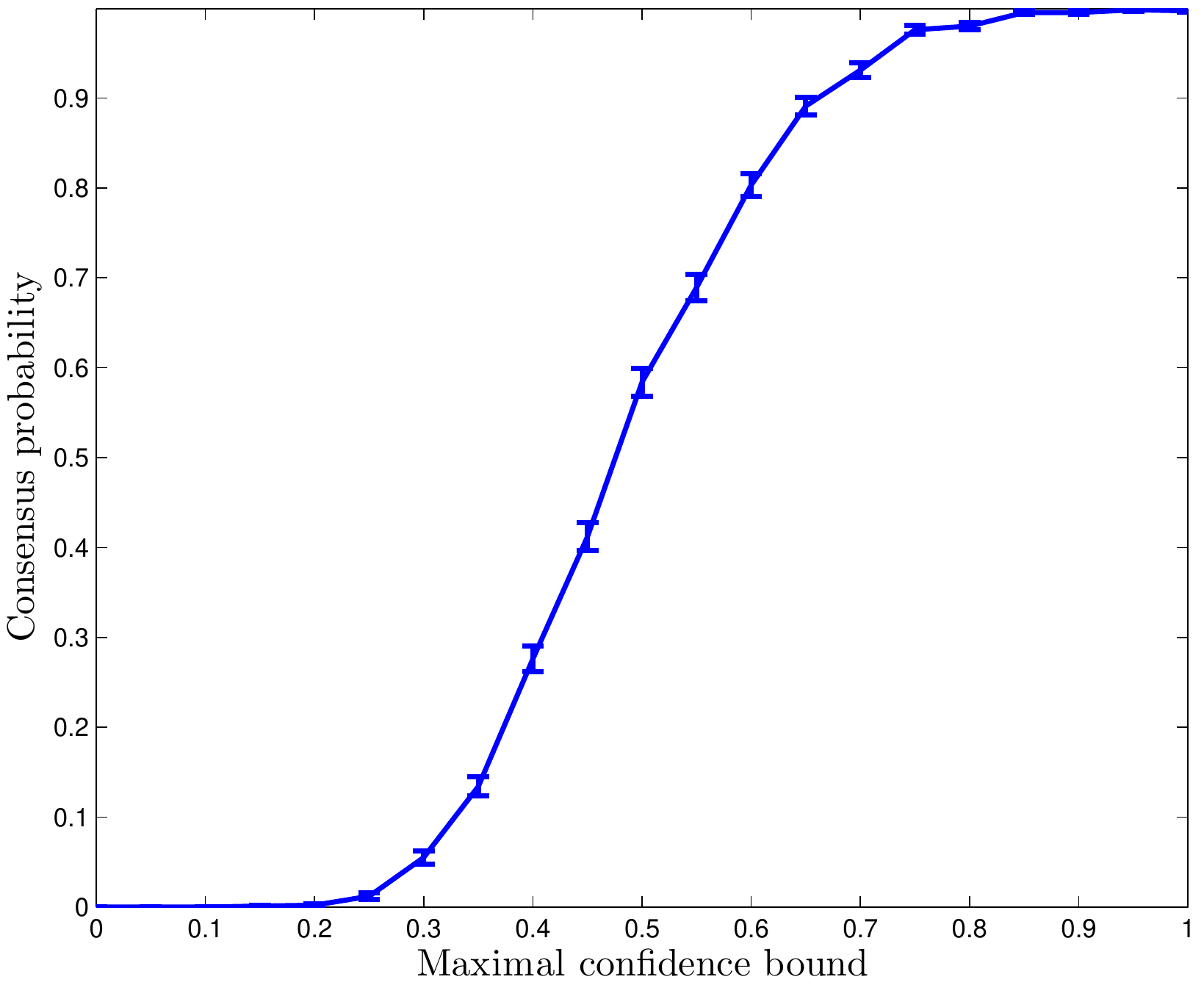}
  \caption{The estimated consensus probability with respect to the maximal confidence bound $r_{\max}$,  where the error bars denote the standard deviations of the estimated probability of reaching consensus at the points of $r_{\max}$.}\label{Fig2}
\end{figure}

Corollary \ref{Cor_2} provides a sufficient and necessary condition for
almost sure consensus when the initial opinions are randomly
distributed. However, for settings when almost sure consensus is not
guaranteed, the probability of achieving consensus is unknown. In the
remainder of this section, we provide simulation results for the consensus
probability of the heterogeneous DW model.
Let $\mu=1/2$ and $n=10$. Suppose
that agent $1$ has a maximal confidence bound $r_{\max}$ whose value is chosen over the set $\{\frac{i}{20}:i=1,2,\ldots,20\}$.
We approximate the consensus probability via the Monte Carlo method. We run 1000 samples for each value of $r_{\max}$.
In each sample, we assume the initial opinions are independently and uniformly distributed on $[0,1]$, while the
confidence bounds of agents $2,3,\ldots,10$ are independently and uniformly
distributed on $[0,r_{\max}]$.  Fig.~\ref{Fig2} shows the estimated consensus
probability of the heterogeneous DW model (\ref{m1})-(\ref{m2}) as a
function of the maximal confidence bound $r_{\max}$.

\section{Proof of convergence results}\label{pomr}

The proof of Theorem \ref{Main_result} requires multiple steps. We adopt
the method of ``transforming the analysis of a stochastic system into the
design of control algorithms" first proposed by \citep{GC:17b}. This method
requires the construction of a new system called as \emph{DW-control
  system} to help with the analysis of the DW model.  The
  following Lemma \ref{robust} gives the connection between the DW model
  and DW-control system.  Also, we introduce a new concept called
  \emph{maximal-confidence cluster} whose basic properties will be provided
  in the following Lemmas \ref{mcc_dis}-\ref{mcc_convex}.  In the following
  Lemmas \ref{diameter0} and \ref{diameter2}, we design control algorithms
  on DW-control system around the maximal-confidence cluster. The proof of
  Theorem~\ref{Main_result} follows from these lemmas.

\subsection{DW-control system and connection to DW model}

Consider the DW protocol (\ref{m1})-(\ref{m2}) where, at each time $t$, the
pair $\{i_t,j_t\}$ is not selected randomly but instead treated as a control
input. In other words, assume that $\{i_t,j_t\}$ is chosen from the set
$\mcN$ arbitrarily as a control signal. We call such a control system the
\emph{DW-control system}.

Given $S\subseteq\real^n$, we say $S$ \emph{is reached at time $t$} if
$x(t)\in S$ and \emph{is reached in the time interval $[t_1,t_2]$} if
there exists $t\in [t_1,t_2]$ such that $x(t)\in S$.

The following lemma builds a connection between the DW model and DW-control system.


\begin{lem}\label{robust}{\textbf{\textup{(Connection between DW model and DW-control system)}}}
  Let $S\subseteq [0,1]^n$ be a set of states. Assume there exists a duration $t^*>0$ such that for any
  $x(0) \in [0,1]^n$, we can find a sequence of pairs
  $\{i_0',j_0'\},\{i_1',j_1'\},\ldots,\{i_{t^*-1}',j_{t^*-1}'\}$ for opinion
  update which guarantees $S$ is reached in the time interval $[0,t^*]$.
  Let $a:=1-\frac{2^{t^*}}{n^{t^*}(n-1)^{t^*}}$. Then, under the DW
  protocol, for any initial state $x(0)\in [0,1]^n$ we have
  \begin{equation*}
    \Prob\left(\tau\geq t\right)\leq a^{\lfloor t/(t^*+1)\rfloor},  \quad \forall t\geq 1,
  \end{equation*}
 where $\tau:=\min\{t': x(t')\in S\}$ is the time when $S$ is firstly reached.
\end{lem}

\begin{pf}
First according to the rule of the DW protocol (\ref{m1})-(\ref{m2}) we get
$x(t)\in [0,1]^n$ for all $t\geq 0$. Also, for any $x(t)\in
  [0,1]^n$, by the assumption of this lemma we can find a sequence of pairs
  $\{i_t',j_t'\},\{i_{t+1}',j_{t+1}'\},\ldots,\{i_{t+t^*-1}',j_{t+t^*-1}'\}$
  for opinion update such that $S$ is reached in $[t,t+t^*]$ under the
  DW-control system.  Thus, under the DW protocol, for any $t\geq 0$ and
$x(t)\in [0,1]^n$ we have
  \begin{equation}\label{robust_1}
    \begin{aligned}
      \Prob&\left(\left\{\mbox{$S$ is reached in $[t,t+t^*]$}\right\}|x(t)\right)\\
      &\qquad\geq \Prob\Big(\bigcap_{s=t}^{t+t^*-1} \big\{\{i_s,j_s\}=\{i_s',j_s'\}\big\}|x(t)\Big)\\
      &\qquad=\prod_{s=t}^{t+t^*-1} \Prob\big(\{i_s,j_s\}=\{i_s',j_s'\}\big)\\
      &\qquad=\frac{1}{|\mcN|^{t^*}}=\frac{2^{t^*}}{n^{t^*}(n-1)^{t^*}},
    \end{aligned}
  \end{equation}
where the first and second equalities use the fact that  $\{i_t,j_t\}$ is uniformly and independently selected from the set
$\mcN$, and  $|\mcN|$ denotes the cardinality of the set $\mcN$.


Set $E_t$ to be the event that $S$ is reached in $[t,t+t^*]$, and
let $E_{t}^c$ be the complement set of $E_{t}$.
For any integer $M>0$ and $x(0)\in[0,1]^n$,
Bayes' Theorem and equation~(\ref{robust_1}) imply
\begin{eqnarray}\label{robust_4}
&&\Prob\big(\left\{\mbox{$S$ is not reached in $[0,(t^*+1)M-1]$}\right\}|x(0)\big)\nonumber\\
&&\enspace=\Prob\Big(\bigcap_{m=0}^{M-1}E_{m(t^*+1)}^c\big|x(0)\Big)\nonumber\\
&&\enspace=\Prob\left(E_{0}^c|x(0)\right)\prod_{m=1}^{M-1}\Prob\Big(E_{m(t^*+1)}^c\big|x(0),\!\!\bigcap_{0\leq m'<m}\!\!E_{m'(t^*+1)}^c\Big)\nonumber\\
&&\enspace\leq \Big(1-\frac{2^{t^*}}{n^{t^*}(n-1)^{t^*}}\Big)^M=a^M.
\end{eqnarray}
For any integer $M>0$ and $x(0)\in[0,1]^n$, by (\ref{robust_4}) we have
\begin{align*}\label{robust_6}
  \Prob\big(\tau & \geq (t^*+1)M |x(0)\big) \nonumber \\
  &=\Prob\big(\left\{\mbox{$S$ is not reached in $[0,(t^*+1)M-1]$}\right\}|x(0)\big)\nonumber \\
  &\leq a^M,
\end{align*}
and, in turn,
\begin{equation*}\label{robust_7}
  \Prob\left(\tau \geq t|x(0) \right)\leq \Prob\left(\tau \geq \Big\lfloor \frac{t}{T} \Big\rfloor T |x(0) \right)\leq a^{\lfloor t/(t^*+1)\rfloor}.
\end{equation*}
  \qquad\hfill \oprocend
\end{pf}

According to Lemma \ref{robust}, to prove the convergence of the DW model,
we only need to design control algorithms for DW-control system such that a
convergence set is reached. Before the design of such control algorithms we
introduce some useful notions.

\subsection{Maximal-confidence clusters and properties}

Recall that we assume $r_1\geq r_2\geq \cdots \geq r_n>0$.  For any opinion
state $x=(x_1,\ldots,x_n)\in [0,1]^n$, let $C_1(x) \subseteq \until{n}$ be
the set of the agents that can connect to agent $1$ directly or indirectly
with the confidence bound $r_1$, i.e., $i \in C_1(x)$ if and only if
$|x_i-x_1|\leq r_1$ or there exists some agents $1',2',\ldots,k' \in
\until{n}$ such that $|x_i-x_{1'}|\leq r_1, |x_{1'}-x_{2'}|\leq r_1,
\ldots, |x_{k'}-x_1|\leq r_1$. From this definition we have $1\in C_1(x)$.

Set $\widetilde{C}_1(x):=\until{n}\setminus C_1(x)$. If
$\widetilde{C}_1(x)$ is not empty, we let $i_2:=\min_{i\in
  \widetilde{C}_1(x)} i$ and define $C_2(x)\subseteq \widetilde{C}_1(x)$ to
be the set of the agents that can connect to agent $i_2$ directly or
indirectly with the confidence bound $r_{i_2}$. Set
$\widetilde{C}_2(x):=\until{n}\setminus (C_1(x) \cup C_2(x))$. If
$\widetilde{C}_2(x)$ is not empty, we let $i_3:=\min_{i\in
  \widetilde{C}_2(x)} i$ and define $C_3(x)\subseteq \widetilde{C}_2(x)$ to
be the set of the agents that can connect to agent $i_3$ directly or
indirectly with the confidence bound $r_{i_3}$. Repeat this process until
there exists an integer $K$ such that $\widetilde{C}_K(x)=\emptyset$. We
call the sets $C_1(x),C_2(x),\ldots,C_K(x)$ \emph{maximal-confidence (MC)
  clusters}. Note that MC clusters are quite different from
connected components in graph theory.

\begin{figure}
  \centering
  \includegraphics[width=3.2in]{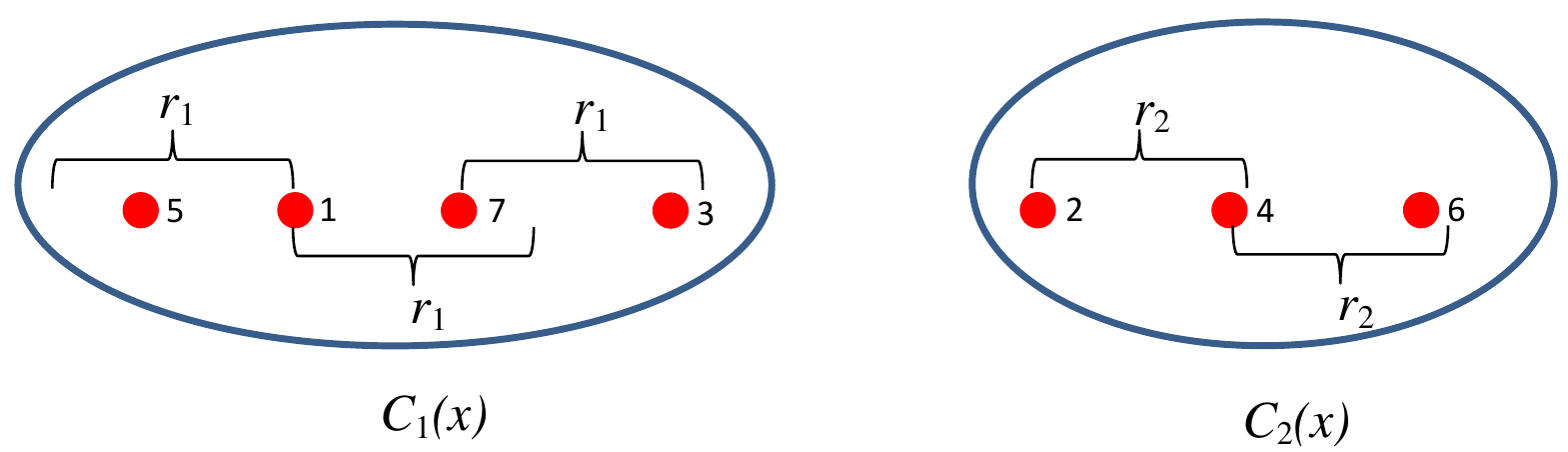}
  \caption{Two MC clusters $C_1(x)$ and $C_2(x)$. The distance between two
    adjacent nodes in $C_1(x)$ (or $C_2(x)$) is not bigger than the $r_1$
    (or $r_2$), but the distance between the agents $3$ and $2$ is bigger
    than $r_1$.}\label{mccluster}
\end{figure}

To illustrate the definition of MC clusters we give an example, visualized
Fig.~\ref{mccluster}: Assume that $n=7$ and that the agents are labeled by
$1,2,\ldots,7$. We suppose $r_1\geq r_2\geq \cdots \geq r_7$. With the
confidence bound $r_1$ the agent $1$ can connect to agents $5$ and $7$, and
the agent $7$ can connect to agent $3$; however agent $3$ cannot connect to
agent $2$. Thus, the first MC cluster $C_1(x)$ is $\{1,3,5,7\}$. The
remaining agents are $2,4,$ and $6$. With the confidence $r_2$ the agent
$2$ can connect to agent $4$, and the agent $4$ can connect to agent $6$,
so the second MC cluster $C_2(x)$ is $\{2,4,6\}$.

The following lemma can be derived immediately from the definition of MC cluster.
\begin{lem}\label{mcc_dis}
  \textbf{\textup{(Distance between maximal-confidence clusters)}} For any
  opinion state $x\in [0,1]^n$ and two different MC clusters $C_{i}(x)$ and
  $C_{j}(x)$, let $r_{\max}^{ij}:=\max_{k\in C_{i}(x)\cup C_{j}(x)}{r_k}$
  be the maximal confidence bound of all agents in $C_{i}(x)$ and
  $C_{j}(x)$.  Then, the opinion values of agents in $C_{i}(x)$ are all
  $r_{\max}^{ij}$ bigger or smaller than those in $C_{j}(x)$, i.e.,
  \begin{align*}
    x_k-x_l&>r_{\max}^{ij}~~~~ \forall k\in C_i(x), l\in C_j(x), \qquad \text{ or } \\
    x_l-x_k&>r_{\max}^{ij}~~~~ \forall k\in C_i(x), l\in C_j(x).
  \end{align*}
\end{lem}


Under the DW protocol (\ref{m1})-(\ref{m2}), the MC
clusters have the convex property as follows.

\begin{lem}\label{mcc_convex}
\textbf{\textup{(Convexity of maximal-confidence clusters)}} Consider the
DW protocol (\ref{m1})-(\ref{m2}) with arbitrary initial state and update
pairs $\{\{i_t,j_t\}\}_{t\geq 0}$.  For any $t\geq 0$ and any
MC cluster $C_{i}(x(t))$, the opinion values of all agents
in $C_{i}(x(t))$ will always stay in the interval $[x_{\min}^i(t),
  x_{\max}^i(t)]$ at the time $s\geq t$, i.e.,
\begin{gather*}
x_{\min}^i(t)\leq x_j(s) \leq x_{\max}^i(t),~~~~\forall j\in C_i(x(t)),  s\geq t,
\end{gather*}
where $x_{\min}^i(t):=\min_{k\in C_{i}(x(t))}{x_k(t)}$ and $x_{\max}^i(t):=\max_{k\in C_{i}(x(t))}{x_k(t)}$ denote the
minimal and maximal opinion values of all agents in $C_{i}(x(t))$ respectively.
\end{lem}

\begin{pf}
Assume that at time $t$ all MC clusters are
$C_1=C_1(x(t)),C_2=C_2(x(t)),\ldots,C_K=C_K(x(t))$.  By Lemma \ref{mcc_dis}
we can order these clusters as
\begin{gather*}\label{mccc_1}
  C_{j_1}\prec C_{j_2} \prec \cdots \prec C_{j_K},
\end{gather*}
and get, for $1\leq k \leq K-1$,
\begin{equation}\label{mccc_2}
  \min_{l\in C_{j_{k+1}}}{x_l(t)}- \max_{l\in C_{j_{k}}}{x_l(t)}  >r^{k,k+1},
\end{equation}
where $C_i\prec C_j$ means that at time $t$ the opinion values of the
agents in $C_i$ are all less than those in $C_j$, and
$r^{k,k+1}:=\max_{l\in C_{j_k}\cup C_{j_{k+1}}}{r_l}$.

By the DW protocol (\ref{m1})-(\ref{m2}), if the update pair $\{i_t,j_t\}$
belongs to different MC clusters then from (\ref{mccc_2})
we have $x_{i_t}(t+1)=x_{i_t}(t)$ and $x_{j_t}(t+1)=x_{j_t}(t)$; if
$\{i_t,j_t\}$ belongs to a same MC cluster $C_{j_k}$ then
$x_{i_t}(t+1)$ and $x_{j_t}(t+1)$ will stay in the interval
$[x_{\min}^{j_{k}}(t),x_{\max}^{j_k}(t)]$.  Thus, for $1\leq k \leq K-1$,
\begin{equation*}
\min_{l\in C_{j_{k+1}}}{x_l(t+1)}- \max_{l\in C_{j_{k}}}{x_l(t+1)} >r^{k,k+1}.
\end{equation*}
Repeating this process yields our result. \oprocend
\end{pf}

With the definition and properties of MC clusters we can
design control algorithms and complete final proof of our results in the
following subsection.

\subsection{Design of control algorithms}

Throughout this subsection we assume $\mu\in[1/2,1)$ and
    $r_1\geq r_2\geq \cdots \geq r_n>0.$ We first design control algorithms
    to split a MC cluster into different MC clusters, or reduce its
    diameter by a certain value in finite time.

\begin{lem}\label{diameter0}
Let $t\geq 0$ and $x(t)\in[0,1]^n$ be arbitrarily given.  Let $C_i(x(t))$
be an arbitrary MC cluster, in which the agents' maximal
and minimal confidence bounds are $r_{\max}^i$ and $r_{\min}^i$
respectively.  Assume
\begin{eqnarray}\label{dia0_00}
  \max_{M,m\in C_i(x(t))} [x_M(t)- x_m(t)]> r_{\min}^i.
\end{eqnarray}
Then, under the DW-control system, there is a sequence of agent pairs
$\{i_{t}',j_{t}'\},\{i_{t+1}',j_{t+1}'\},\ldots,\{i_{t+t^*-1}',j_{t+t^*-1}'\}$
with $$t^*\leq (|C_i(x(t))|-1)^2 \left(1+\lceil\log_{1-\mu}
  r_{\min}^i/ r_{\max}^i\rceil \right)$$ for opinion update, such that
one of the following two results holds:
\begin{enumerate}
\item the agents in $C_i(x(t))$ split into different MC clusters at time $t+t^*$; and
\item we have
  \begin{multline*}
    \max_{M,m\in C_i(x(t))}  [x_M(t+t^*)- x_m(t+t^*)]\\
    \leq \max_{M,m\in C_i(x(t))} [x_M(t)- x_m(t)]-(1-\mu)^2 r_{\min}^i.
  \end{multline*}
\end{enumerate}
\end{lem}
The proof of Lemma \ref{diameter0} is quite complicated. We put it in Appendix \ref{APP_0}.

\begin{rem}\label{rmk_1}
  The result in Lemma \ref{diameter0} cannot be extended to the case when
  $\mu<1/2$. For example, assume $n=3$, $(r_1,r_2,r_3)=(0.4,0.3,0.2)$, and
  $(x_1(0),x_2(0),x_3(0))=(0.1-\varepsilon,0.1+\varepsilon,0.5)$, where
  $\varepsilon\in(0,0.2(1-\mu)^2)$ is a small constant. Then, the three
  agents form a MC cluster, however the interaction exists only between
  agents $1$ and $2$. Because
  \begin{equation*}
    \lim\nolimits_{t\to\infty}
    \begin{bmatrix}
      1-\mu & \mu  \\
      \mu & 1-\mu
    \end{bmatrix}^t =
    \begin{bmatrix}
      0.5 & 0.5  \\
      0.5 & 0.5
    \end{bmatrix},
  \end{equation*}
  we know $x_1(t)\uparrow 0.1$, $x_2(t)\downarrow 0.1$ as $t\to\infty$ if
  we always choose $\{1,2\}$ as the opinion update pair. In fact, we cannot
  find a finite sequence of opinion update pairs such that either result
  (i) or result (ii) in Lemma \ref{diameter0} holds.
\end{rem}

For any opinion state $x\in[0,1]^n$ and any MC cluster
$C_i(x)$, we say that $C_i(x)$ is a \emph{complete cluster} if any agent in
$C_i(x)$ can interact with others with the minimal confidence bound of
$C_i(x)$, i.e.,
\begin{equation*}
\max_{j,k\in C_i(x)} |x_j-x_k|\leq \min_{j\in C_i(x)} r_j.
\end{equation*}
 Lemma \ref{diameter0} leads to control algorithms such that
  all MC clusters become complete clusters in finite time.


\begin{lem}\label{diameter2}
Consider the DW-control system. Then for any initial
state, there exists a sequence of agent pairs
$\{i_0',j_0'\},\{i_{1}',j_{1}'\},\ldots,\{i_{T-1}',j_{T-1}'\}$ with
\begin{equation*}\label{dia2_00}
 T \leq (n-1)^2 \left(1+\left\lceil\log_{1-\mu} \frac{r_n}{r_1}\right\rceil \right)\left\lceil \frac{1-r_{n}}{(1-\mu)^2 r_{n}} \right\rceil
\end{equation*}
for opinion update such that all MC clusters are complete clusters at time $T$.
\end{lem}
\begin{pf}
Assume that at time $t$ all agents are divided into  $K_t$ MC clusters labeled as $C_1(x(t))$,$\ldots,C_{K_t}(x(t))$.  Define
\begin{multline*}
f_i(t):=
\left\{
\begin{aligned}
& 0,~  \mbox{if $C_i(x(t))$ is a complete cluster},\\
& \max_{M,m\in C_i(x(t))} [x_M(t)- x_m(t)],~\mbox{otherwise},
\end{aligned}\right.
\end{multline*}
and $F(t):=\sum_{i=1}^{K_t} f_i(t)$. By Lemma \ref{mcc_dis} we have $F(t)\in \{0\}\cup (r_n,1]$,  and all MC clusters are complete clusters at time $t$ if and only if $F(t)=0$. If $F(t)>0$, by Lemmas \ref{mcc_dis} and \ref{diameter0} there is a sequence of agent pairs
$\{i_{t}',j_{t}'\},\{i_{t+1}',j_{t+1}'\},\ldots,\{i_{t+t^*-1}',j_{t+t^*-1}'\}$
with $$t^*\leq (n-1)^2 \left(1+\lceil\log_{1-\mu} r_{n}/ r_{1}\rceil \right)$$ for opinion update, such that
$$F(t+t^*)\leq F(t)-(1-\mu)^2 r_n.$$
With this process repeated, we can find a sequence of agent pairs
$\{i_{0}',j_{0}'\},\{i_{1}',j_{1}'\},\ldots,\{i_{T-1}',j_{T-1}'\}$
with $$T \leq (n-1)^2 \left(1+\left\lceil\log_{1-\mu} \frac{r_{n}}{r_{1}}\right\rceil \right)\left\lceil \frac{1-r_{n}}{(1-\mu)^2 r_{n}} \right\rceil$$ for opinion update, such that $F(T)=0$.
\oprocend
\end{pf}

\subsection{Final proofs}

\begin{pf*}{Proof of Theorem~\ref{Main_result}}
The proof of convergence rate partly uses the idea appearing in Section II.B of \citep{SB-AG-BP-DS:06}.
Let $\tau$ be the first time when all MC clusters are complete clusters under the
DW protocol (\ref{m1})-(\ref{m2}).  By Lemmas ~\ref{diameter2} and \ref{robust},
\begin{gather}\label{Conr_1}
\Prob\left(\tau\geq t\right)\leq c^{\lfloor t/(T+1)\rfloor}, ~~~~\forall t\geq 1.
\end{gather}
Then $\Prob(\tau<\infty)=1$.
Label the MC clusters as $C_1,\ldots,C_K$ at time $\tau$. By Lemmas \ref{mcc_dis} and \ref{mcc_convex},
for $t\geq\tau$  all MC clusters $C_1,\ldots,C_K$ remain unchanged, i.e., if node $i$ belongs to a cluster $C_j$ at time $\tau$ then it will always belong to $C_j$ for $t>\tau$.

Next, we consider the dynamics when $t\geq \tau$.
 Define the matrix $P(t)\in[0,1]^{n\times n}$ by
\begin{multline}\label{Conr_2}
(P_{ii}(t),P_{jj}(t),P_{ij}(t),P_{ji}(t)):= \\
\left\{
\begin{aligned}
  (1-\mu,1-\mu,\mu,\mu),~~ & \mbox{if $\{i,j\}$ is the opinion update}\\
 & \mbox{pair at time $t$ and belongs the}\\
 & \mbox{same MC cluster,}\\
  (1,1,0,0),~~~~ & \mbox{otherwise.}
\end{aligned}\right.
\end{multline}
for all $i<j$. Then $P(t)$ is a symmetric stochastic matrix.
By the protocol (\ref{m1})-(\ref{m2}), and the facts that
  all MC clusters are complete clusters, and two agents in different MC clusters have no interaction,  we can get
\begin{equation*}
x(t+1)=P(t)x(t),~~\forall t\geq\tau.
\end{equation*}
Also, because $C_1,\ldots,C_K$ remain unchanged,
there exists a permutation matrix $Q\in\{0,1\}^{n\times n}$ such that
\begin{eqnarray}\label{Conr_3}
  Q^\top P(t) Q=\mbox{diag}\left(W_1(t),\ldots,W_K(t)\right):=W(t),
\end{eqnarray}
where $W_k(t)$ is a $|C_k|\times |C_k|$ matrix corresponding to the MC cluster $C_k$.
For $z(t):=Q^\top x(t)$,  we have
\begin{align}
  z(t+1) &=Q^\top x(t+1)=Q^\top P(t)x(t) \nonumber\\
  &=Q^\top P(t) Q z(t)=W(t)z(t)\nonumber\\
  &=W(t)\cdots W(\tau) z(\tau)\nonumber\\
  &=\mbox{diag}\big(W_1(t)\cdots W_1(\tau),\ldots,\nonumber\\
  &~~~~~~~~~~~W_K(t)\cdots W_K(\tau)\big) z(\tau).
  \label{Conr_4}
\end{align}
Set $J_0:=0$ and $J_k:=|C_1|+|C_2|+\cdots+|C_k|$ for $1\leq k\leq K$. Let
\begin{gather*}\label{Conr_5}
\vec{z}_k(t):=(z_{J_{k-1}+1}(t),z_{J_{k-1}+2}(t)\ldots,z_{J_k}(t))^\top.
\end{gather*}
By (\ref{Conr_4}), for $1\leq k\leq K$ we have
\begin{gather}\label{Conr_6}
  \vec{z}_k(t+1)=W_k(t)\vec{z}_k(t)=W_k(t)\cdots W_k(\tau) \vec{z}_k(\tau).
\end{gather}
Let $\overline{z_k}(t):=\frac{\mathbf{1}_{|C_k|}^\top \vec{z}_k(t) }{|C_k|} \mathbf{1}_{|C_k|}$ be the average vector of $\vec{z}_k(t)$, where $\mathbf{1}_{|C_k|}=(1,\ldots,1)^\top$ is a $|C_k|-$dimensional column vector.
By (\ref{Conr_2}) and
(\ref{Conr_3}) we know $W_k(t)$ is a  symmetric  stochastic matrix so that
\begin{eqnarray}\label{Conr_7}
\overline{z_k}(t+1)&&=\frac{[\mathbf{1}_{|C_k|}^\top W_k(t)] \vec{z}_k(t) }{|C_k|} \mathbf{1}_{|C_k|}=\frac{\mathbf{1}_{|C_k|}^\top \vec{z}_k(t) }{|C_k|} \mathbf{1}_{|C_k|}\nonumber\\
&&=\overline{z_k}(t)=\cdots=\overline{z_k}(\tau).
\end{eqnarray}
Set $\vec{y}_k(t):=\vec{z}_k(t)-\overline{z_k}(t)$. We note that if $|C_k|=1$ then $\vec{y}_k(t)=0$. Thus, we only need to consider the case when
$|C_k|\geq 2$.
By (\ref{Conr_6}) and  (\ref{Conr_7}) we have
\begin{eqnarray*}\label{Conr_8}
  \vec{y}_k(t+1)&&=\vec{z}_k(t+1)-\overline{z_k}(t+1)\\
 && =W_k(t) (\vec{z}_k(t)-\overline{z_k}(t))=W_k(t) \vec{y}_k(t).
\end{eqnarray*}
Therefore, we can write
\begin{multline}\label{Conr_9}
 \E\left[ \|\vec{y}_k(t+1)\|^2 |\vec{y}_k(t) \right]
 = \vec{y}_k^{\top}(t) \E \left[W_k^2(t)\right] \vec{y}_k(t).
\end{multline}
Because an agent pair for  opinion update
is selected uniformly and independently from $\mcN$ at each time, by (\ref{Conr_2}) we have
\begin{align*} 
  \E \left[W_k^2(t)\right]_{ij} &=\sum_{l=1}^{|C_k|} \E [W_k(t)]_{il} [W_k(t)]_{lj}\\
  &=\E \left([W_k(t)]_{ii} [W_k(t)]_{ij}+[W_k(t)]_{ij} [W_k(t)]_{jj}\right)\\
  &=\Prob\left( [W_k(t)]_{ij}>0 \right)\left[(1-\mu)\mu+\mu(1-\mu)\right]\\
  &=\frac{4\mu(1-\mu)}{n(n-1)}, ~~~~~~~~\forall i\neq j,
\end{align*}
and then $\ds   \E \left[W_k^2(t) \right]_{ii}=1-\frac{4\mu(1-\mu)(|C_k|-1)}{n(n-1)}$.
Let $1=\lambda_1\geq \lambda_2\geq\cdots\geq\lambda_{|C_k|}$ be the
eigenvalues of $\E W_k^2(t)$, while $\xi_1,\ldots,\xi_{|C_k|}$ be the
corresponding unit right eigenvectors.  It can be computed that
$$\lambda_2=\cdots=\lambda_{|C_k|}=1-\frac{|C_k|4\mu(1-\mu)}{n(n-1)}.$$
Also, $\xi_1=\frac{\mathbf{1}_{|C_k|}}{\sqrt{|C_k|}}\perp \vec{y}_k(t)$,
and $\xi_i \perp \xi_j$ for $i\neq j$, we have
\begin{eqnarray*}\label{Conr_12}
&&\vec{y}_k^{\top}(t) \E \left[W_k^2(t)\right] \vec{y}_k(t)\\
&&~~=\left(\sum_{l=2}^n (\xi_l^\top \vec{y}_k(t))\xi_l \right)^\top \left(\sum_{l=2}^n (\xi_l^{\top} \vec{y}_k(t))\lambda_l \xi_l \right)\\
&&~~=\left(1-\frac{|C_k|4\mu(1-\mu)}{n(n-1)}\right) \vec{y}_k^{\top}(t) \vec{y}_k(t).
\end{eqnarray*}
Combining this with (\ref{Conr_9}) we have
\begin{multline}\label{Conr_13}
 \E\|\vec{y}_k(t+1)\|^2  = \left(1-\frac{|C_k|4\mu(1-\mu)}{n(n-1)}\right) \E \|\vec{y}_k(t)\|^2.
\end{multline}
Using (\ref{Conr_13}) repeatedly we can get
\begin{align}\label{Conr_14}
\E &\left[\sum_{k=1}^K \|\vec{y}_k(t)\|^2 |t\geq \tau,C_1,\ldots,C_K\right] \nonumber\\
&= \E \bigg[\sum_{1\leq k\leq K, |C_k|\geq 2} \left(1-\frac{|C_k|4\mu(1-\mu)}{n(n-1)}\right)^{t-\tau}    \nonumber\\
&~~~~~~\qquad\times \|\vec{y}_k(\tau)\|^2 \big|\,t\geq \tau,C_1,\ldots,C_K\bigg] \nonumber\\
&\leq \frac{n}{4} \left(1-\frac{8\mu(1-\mu)}{n(n-1)}\right)^{t-\tau},
\end{align}
where the inequality uses the Popoviciu inequality \citep{TP:35} which says for any real numbers $b_1,\ldots,b_m$ we have
$$\frac{1}{m}\sum_{l=1}^m \left(b_l-\frac{b_1+\cdots+b_m}{m}\right)^2 \leq \frac{1}{4} \left(\max_l b_l-\min_l b_l \right)^2.$$
Let $x^*:=Q (\overline{z_1}^\top (\tau),\ldots,\overline{z_K}^{\top}(\tau))^{\top}$.
Since $z(t)$ is a rearrangement of entries of $x(t)$, by (\ref{Conr_1}), (\ref{Conr_14})
and the total probability formula we have
\begin{align}\label{Conr_15}
\E&\|x(t)-x^*\|^2 \nonumber\\
&=\Prob\left(\tau>\frac{t}{2}\right) \E\left[ \|x(t)-x^*\|^2 \big|\tau>\frac{t}{2} \right] \nonumber\\
&\qquad+\Prob\left(\tau\leq\frac{t}{2}\right) \E\left[\|x(t)-x^*\|^2 \big|\tau\leq\frac{t}{2} \right] \nonumber\\
&\leq n c^{\lfloor \frac{t}{2(T+1)}\rfloor} + \frac{n}{4} \left(1-\frac{8\mu(1-\mu)}{n(n-1)}\right)^{\lfloor \frac{t}{2}\rfloor}.
\end{align}
For any constant $\varepsilon>0$,  by (\ref{Conr_15}) and the Markov's inequality we can get
\begin{eqnarray*}\label{Conr_16}
\sum_{t=1}^{\infty} \Prob\left(\|x(t)-x^*\|>\varepsilon\right)\leq \sum_{t=1}^{\infty} \frac{\E \|x(t)-x^*\|^2}{\varepsilon^2}<\infty,
\end{eqnarray*}
then by the Borel-Cantelli lemma we have a.s. $x(t)\rightarrow x^*$ as $t\rightarrow\infty$.
By Lemma \ref{mcc_dis} and the definition of $x^*$ we obtain $x_i^*=x_j^*$ or $|x_i^*-x_j^*|>\max\{r_i,r_j\}$ for any $i\neq j$.
~~\oprocend
\end{pf*}

\begin{pf*}{Proof of Corollary~\ref{Cor_1}}
By Theorem \ref{Main_result} we have $x(t)$ a.s.   converges to a limit point $x^*\in [0,1]^n$ which satisfies
either $|x_1^*-x_i^*|=0$ or $|x_1^*-x_i^*|>r_1$ for all $2\leq i\leq n$.  Because $r_1\geq 1$, we have $|x_1^*-x_i^*|=0$ for all $2\leq i\leq n$,
which indicates $x^*$ is a consensus state.
\oprocend
\end{pf*}

\begin{pf*}{Proof of Corollary~\ref{Cor_2}}
If $r_1\geq 1$, then Corollary~\ref{Cor_1} implies that the system reaches
consensus a.s.

If $r_1<1$, then equation~(\ref{cor2_0}) implies
\begin{eqnarray*}\label{cor2_1}
  \Prob\Big(x_1(0)\in \Big[0,\frac{1-r_1}{3}\Big], \bigcap_{i=2}^n \Big\{x_i(0)\in \Big[\frac{2+r_1}{3},1\Big]\Big\} \Big)\\
  \geq \rhomin \left(\frac{1-r_1}{3}\right)^n .
\end{eqnarray*}
Also, if $x_1(0)\in[0,\frac{1-r_1}{3}]$ and the event $\bigcap_{i=2}^n
\{x_i(0)\in [\frac{2+r_1}{3},1]\}$ takes place, then
$|x_1(0)-x_i(0)|=\frac{1+2 r_1}{3}>r_1$ for $2\leq i\leq n$. In turn, this
implies that the system cannot reach consensus because the agent $1$ can
never interact with the agents $2,\ldots,n$.  \oprocend
\end{pf*}

\section{Conclusions}\label{Conclusions}

Bounded confidence (BC) models of opinion dynamics adopt a mechanism
whereby individuals are not willing to accept other opinions if these other
opinions are beyond a certain confidence bound. These models have attracted
significant mathematical and sociological attention in recent years.  One
well-known BC model is the Deffuant-Weisbuch (DW) model, in which a pair of
agents is selected randomly at each time step, and each agent in the pair
updates its opinion if the other agent's opinion in the pair is within its
confidence bound.  Because the inter-agent topology of the DW model is
coupled with the agents' states, the heterogeneous DW model is hard to
analyze. This paper proves the convergence of a heterogeneous DW model and
shows the mean-square error is bounded by a negative exponential function of time.

As directions for future research, it remains to prove the convergence of
the heterogeneous DW model with the weighting factor
  $\mu\in(0,1/2)$. From Remark \ref{rmk_1}, the convergence for the case
  $\mu\in(0,1/2)$ cannot be deduced directly by the current method.  A more
  ingenious control design may be required to establish that the DW-control
  system converges to a set with invariant topology in finite time.


\begin{ack}
The authors thank Professors Jiangbo Zhang and Yiguang Hong for their kind
advice and candid clarification about previous works.
\end{ack}

\bibliographystyle{plainnat}
\bibliography{alias,FB,Main}

\begin{thebibliography}{25}
\providecommand{\natexlab}[1]{#1}
\providecommand{\url}[1]{\texttt{#1}}
\expandafter\ifx\csname urlstyle\endcsname\relax
  \providecommand{\doi}[1]{doi: #1}\else
  \providecommand{\doi}{doi: \begingroup \urlstyle{rm}\Url}\fi

\bibitem[Boyd et~al.(2006)Boyd, Ghosh, Prabhakar, and Shah]{SB-AG-BP-DS:06}
S.~Boyd, A.~Ghosh, B.~Prabhakar, and D.~Shah.
\newblock Randomized gossip algorithms.
\newblock \emph{IEEE Transactions on Information Theory}, 52\penalty0
  (6):\penalty0 2508--2530, 2006.
\newblock \doi{10.1109/TIT.2006.874516}.

\bibitem[Bullo(2019)]{FB:19}
F.~Bullo.
\newblock \emph{Lectures on Network Systems}.
\newblock Kindle Direct Publishing, {1.3} edition, July 2019.
\newblock ISBN 978-1986425643.
\newblock URL \url{http://motion.me.ucsb.edu/book-lns}.
\newblock With contributions by J. Cort{\'e}s, F. D\"orfler, and S.
  Mart{\'\i}nez.

\bibitem[Chazelle and Wang(2017)]{BC-CW:17}
B.~Chazelle and C.~Wang.
\newblock Inertial {Hegselmann-Krause} systems.
\newblock \emph{IEEE Transactions on Automatic Control}, 62\penalty0
  (8):\penalty0 3905--3913, 2017.
\newblock \doi{10.1109/TAC.2016.2644266}.

\bibitem[Chen(2017)]{GC:17b}
G.~Chen.
\newblock Small noise may diversify collective motion in {Vicsek} model.
\newblock \emph{IEEE Transactions on Automatic Control}, 62\penalty0
  (2):\penalty0 636--651, 2017.
\newblock \doi{10.1109/TAC.2016.2560144}.

\bibitem[Como and Fagnani(2011)]{GC-FF:11}
G.~Como and F.~Fagnani.
\newblock Scaling limits for continuous opinion dynamics systems.
\newblock \emph{Annals of Applied Probability}, 21\penalty0 (4):\penalty0
  1537--1567, 2011.
\newblock \doi{10.1214/10-AAP739}.

\bibitem[Deffuant et~al.(2000)Deffuant, Neau, Amblard, and
  Weisbuch]{GD-DN-FA-GW:00}
G.~Deffuant, D.~Neau, F.~Amblard, and G.~Weisbuch.
\newblock Mixing beliefs among interacting agents.
\newblock \emph{Advances in Complex Systems}, 3\penalty0 (1/4):\penalty0
  87--98, 2000.
\newblock \doi{10.1142/S0219525900000078}.

\bibitem[DeGroot(1974)]{MHDG:74}
M.~H. DeGroot.
\newblock Reaching a consensus.
\newblock \emph{Journal of the American Statistical Association}, 69\penalty0
  (345):\penalty0 118--121, 1974.
\newblock \doi{10.1080/01621459.1974.10480137}.

\bibitem[{French~Jr.}(1956)]{JRPF:56}
J.~R.~P. {French~Jr.}
\newblock A formal theory of social power.
\newblock \emph{Psychological Review}, 63\penalty0 (3):\penalty0 181--194,
  1956.
\newblock \doi{10.1037/h0046123}.

\bibitem[Friedkin(1998)]{NEF:98}
N.~E. Friedkin.
\newblock \emph{A Structural Theory of Social Influence}.
\newblock Cambridge University Press, 1998.
\newblock ISBN 9780521454827.

\bibitem[Friedkin and Johnsen(1990)]{NEF-ECJ:90}
N.~E. Friedkin and E.~C. Johnsen.
\newblock Social influence and opinions.
\newblock \emph{Journal of Mathematical Sociology}, 15\penalty0 (3-4):\penalty0
  193--206, 1990.
\newblock \doi{10.1080/0022250X.1990.9990069}.

\bibitem[Harary(1959)]{FH:59}
F.~Harary.
\newblock A criterion for unanimity in {F}rench's theory of social power.
\newblock In D.~Cartwright, editor, \emph{Studies in Social Power}, pages
  168--182. University of Michigan, 1959.
\newblock ISBN 0879442301.
\newblock URL \url{http://psycnet.apa.org/psycinfo/1960-06701-006}.

\bibitem[Hegselmann and Krause(2002)]{RH-UK:02}
R.~Hegselmann and U.~Krause.
\newblock Opinion dynamics and bounded confidence models, analysis, and
  simulations.
\newblock \emph{Journal of Artificial Societies and Social Simulation},
  5\penalty0 (3), 2002.
\newblock URL \url{http://jasss.soc.surrey.ac.uk/5/3/2.html}.

\bibitem[Katz and Lazarsfeld(1955)]{EK-PFL:55}
E.~Katz and P.~F. Lazarsfeld.
\newblock \emph{Personal Influence: {T}he Part Played by People in the Flow of
  Mass Communications}.
\newblock Free Press, 1955.
\newblock ISBN 9781412805070.

\bibitem[Latan\'{e}(1981)]{BL:81}
B.~Latan\'{e}.
\newblock The psychology of social impact.
\newblock \emph{American Psychologist}, 36\penalty0 (4):\penalty0 343--365,
  1981.
\newblock \doi{10.1037/0003-066X.36.4.343}.

\bibitem[Latan\'{e}(1996)]{BL:96}
B.~Latan\'{e}.
\newblock Dynamic social impact: {T}he creation of culture by communication.
\newblock \emph{Journal of Communication}, 46\penalty0 (4):\penalty0 13--25,
  1996.
\newblock \doi{10.1111/j.1460-2466.1996.tb01501.x}.

\bibitem[Lorenz(2005)]{JL:05a}
J.~Lorenz.
\newblock A stabilization theorem for dynamics of continuous opinions.
\newblock \emph{Physica A: Statistical Mechanics and its Applications},
  355\penalty0 (1):\penalty0 217--223, 2005.
\newblock \doi{10.1016/j.physa.2005.02.086}.

\bibitem[Lorenz(2007)]{JL:07b}
J.~Lorenz.
\newblock Continuous opinion dynamics under bounded confidence: {A} survey.
\newblock \emph{International Journal of Modern Physics~C}, 18\penalty0
  (12):\penalty0 1819--1838, 2007.
\newblock \doi{10.1142/S0129183107011789}.

\bibitem[Lorenz(2010)]{JL:09}
J.~Lorenz.
\newblock Heterogeneous bounds of confidence: {M}eet, discuss and find
  consensus!
\newblock \emph{Complexity}, 4\penalty0 (15):\penalty0 43--52, 2010.
\newblock \doi{10.1002/cplx.20295}.

\bibitem[MirTabatabaei and Bullo(2012)]{AM-FB:11f}
A.~MirTabatabaei and F.~Bullo.
\newblock Opinion dynamics in heterogeneous networks: {C}onvergence conjectures
  and theorems.
\newblock \emph{SIAM Journal on Control and Optimization}, 50\penalty0
  (5):\penalty0 2763--2785, 2012.
\newblock \doi{10.1137/11082751X}.

\bibitem[Popoviciu(1935)]{TP:35}
T.~Popoviciu.
\newblock Sur les equations algebriques ayant toutes leurs racines reelles.
\newblock \emph{Mathematica}, 9:\penalty0 129--145, 1935.

\bibitem[Proskurnikov and Tempo(2017)]{AVP-RT:17}
A.~V. Proskurnikov and R.~Tempo.
\newblock A tutorial on modeling and analysis of dynamic social networks. {Part
  I}.
\newblock \emph{Annual Reviews in Control}, 43:\penalty0 65--79, 2017.
\newblock \doi{10.1016/j.arcontrol.2017.03.002}.

\bibitem[Proskurnikov and Tempo(2018)]{AVP-RT:18}
A.~V. Proskurnikov and R.~Tempo.
\newblock A tutorial on modeling and analysis of dynamic social networks. {Part
  II}.
\newblock \emph{Annual Reviews in Control}, 45:\penalty0 166--190, 2018.
\newblock \doi{10.1016/j.arcontrol.2018.03.005}.

\bibitem[Weisbuch et~al.(2002)Weisbuch, Deffuant, Amblard, and
  Nadal]{GW-GD-FA-JN:02}
G.~Weisbuch, G.~Deffuant, F.~Amblard, and J.~P. Nadal.
\newblock Meet, discuss, and segregate!
\newblock \emph{Complexity}, 7\penalty0 (3):\penalty0 55--63, 2002.
\newblock \doi{10.1002/cplx.10031}.

\bibitem[Zhang and Chen(2015)]{JZ-GC:15}
J.~Zhang and G.~Chen.
\newblock Convergence rate of the asymmetric {Deffuant-Weisbuch} dynamics.
\newblock \emph{Journal of Systems Science and Complexity}, 28\penalty0
  (4):\penalty0 773--787, 2015.
\newblock \doi{10.1007/s11424-015-3240-z}.

\bibitem[Zhang and Hong(2013)]{JZ-YH:13}
J.~Zhang and Y.~Hong.
\newblock Opinion evolution analysis for short-range and long-range
  {Deffuant-Weisbuch} models.
\newblock \emph{Physica A: Statistical Mechanics and its Applications},
  392\penalty0 (21):\penalty0 5289--5297, 2013.
\newblock \doi{10.1016/j.physa.2013.07.014}.

\end{thebibliography}

\appendix
\section{The proof of Lemma \ref{diameter0}}\label{APP_0}
The proof of this lemma is identical for all cases $t=0,1,2,\ldots$.  To
simplify the exposition we consider only the case when $t=0$.

Assume the agents $j$ and $k$ have the minimal and maximal opinions among
$C_i(x(0))$ at time $0$ respectively, i.e.,
\begin{eqnarray*}
x_j(0)= \min_{m\in C_i(x(0))} x_m(0), ~~ x_k(0)= \max_{m\in C_i(x(0))} x_m(0).
\end{eqnarray*}
Also, assume that the agent $l$  has the maximal confidence bound $r_{\max}^i$ in $C_i(x(0))$.

We first consider the case when  $x_l(0)\geq \frac{x_k(0)+x_j(0)}{2}$.
 From (\ref{dia0_00}) we have
\begin{eqnarray}\label{dia0_03}
x_l(0) \geq x_j(0)+ r_{\min}^i/2.
\end{eqnarray}
Let $$\underline{A}(s):=\{m\in C_i(x(0)): x_m(s)< x_j(0)+(1-\mu)^2 r_{\min}^i\}.$$
We aim to control the agent $l$ to episodically come and pull
out one more agent from $\underline{A}(s)$, or otherwise we have a split of clusters.
 The control strategy can
be divided into the following steps:

 Step $1$:  Control
 the agent pairs for opinion update until one of the following two events happens:\\
({\bf{E1}}) The agents in $C_i(x(0))$ split into different MC clusters;\\
({\bf{E2}}) $|\underline{A}(s)|=|\underline{A}(0)|-1$, where $|\cdot|$ denote the cardinality of a set.

Let $i_0'$ be the agent in $C_i(x(0))$ which has the smallest opinion within the confidence bound of agent $l$ at time $0$, i.e.,
\begin{eqnarray*}\label{dia0_04}
&&i_0'=\mathop{\arg\min}_{m\in C_i(x(0))} \{x_m(0):|x_l(0)-x_m(0)|\leq r_l \}.
\end{eqnarray*}
\begin{figure}
  \centering
  \includegraphics[width=2in]{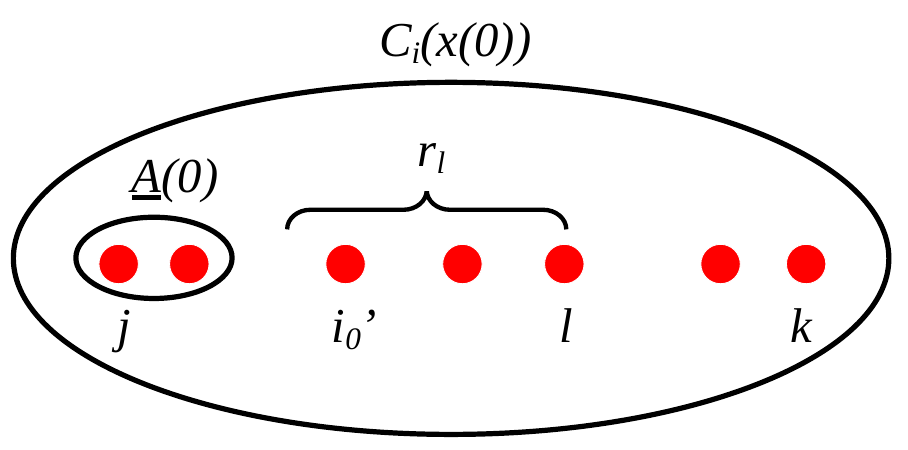}
  \caption{An example for the relation of $C_i(x(0))$, $\underline{A}(0)$, and agents $j$, $k$, $l$, and $i_0'$.}\label{lemma8fig}
\end{figure}
An example for the relation of  $C_i(x(0))$, $\underline{A}(0)$, and agents $j$, $k$, $l$, and $i_0'$ is shown in Fig. \ref{lemma8fig}.
Set $$T_1:=\max\left\{\left\lceil\log_{1-\mu} \frac{r_{i_0'}}{x_l(0)-x_{i_0'}(0)}\right\rceil,0\right\}.$$
 We can get $T_1\leq  \lceil\log_{1-\mu} r_{n} \rceil$ is uniformly bounded.
Choose $\{i_0',l\}$ as the agent pair for opinion update at times  $0,1,\ldots,T_1$.
If $T_1=0$, we have $x_l(0)-x_{i_0'}(0)\leq r_{i_0'}$,
then by the protocol (\ref{m1})-(\ref{m2}) and the fact of $\mu\in[1/2,1)$ we get
\begin{multline}\label{dia0_05}
 x_{l}(1)=(1-\mu)  x_l(0)+ \mu x_{i_0'}(0)\\
 \leq (1-\mu)  x_{i_0'}(0)+ \mu x_{l}(0) =  x_{i_0'}(1)
\end{multline}
If $T_1\geq 1$, by the definition of $T_1$ we have
\begin{eqnarray}\label{dia0_06}
&&T_1-1<\log_{1-\mu} \frac{r_{i_0'}}{x_l(0)-x_{i_0'}(0)} \leq T_1 \iff  \nonumber\\
&&\enspace (1-\mu)^{-T_1+1} r_{i_0'}< x_l(0)-x_{i_0'}(0) \leq (1-\mu)^{-T_1} r_{i_0'}.\nonumber\\
\end{eqnarray}
Using (\ref{dia0_06}) and the protocol
(\ref{m1})-(\ref{m2}) repeatedly we obtain
\begin{eqnarray*}\label{dia0_07}
\left\{
\begin{array}{ll}
x_{i_0'}(s)=x_{i_0'}(0) \\
x_l(s)=x_{i_0'}(0)+(1-\mu)^s (x_l(0)-x_{i_0'}(0))
\end{array}
\right.
\end{eqnarray*}
for $s=1,\ldots,T_1$,
and
\begin{eqnarray}\label{dia0_08}
\left\{
\begin{array}{ll}
x_{i_0'}(T_1+1)=x_{i_0'}(0)+\mu(1-\mu)^{T_1}(x_l(0)-x_{i_0'}(0)) \nonumber\\
x_l(T_1+1)=x_{i_0'}(0)+(1-\mu)^{T_1+1}(x_l(0)-x_{i_0'}(0)) \nonumber
\end{array}
\right..\\
\end{eqnarray}
We continue our discussion by considering the following two cases:

\emph{Case I}: $i_0'\in \underline{A}(0)$.
By (\ref{dia0_05}), (\ref{dia0_06}), and (\ref{dia0_08}) we get
\begin{eqnarray}\label{dia0_08_1}
    x_{i_0'}(T_1+1) &&\geq x_l(T_1+1)\nonumber\\
    &&=x_{i_0'}(0)+(1-\mu)^{T_1+1}(x_l(0)-x_{i_0'}(0))\nonumber\\
    &&>x_{i_0'}(0)+(1-\mu)^2 r_{i_0'}\nonumber\\
    &&\geq x_{j}(0)+ (1-\mu)^2 r_{\min}^i.
\end{eqnarray}
Because all agents except $l$ and $i_0'$ keep their opinions invariant during the time $[0,T_1+1]$,
by (\ref{dia0_08_1}) we have
\begin{equation}\label{dia0_08_2}
|\underline{A}(T_1+1)|=|\underline{A}(0)|-1.
\end{equation}

\emph{Case II}: $i_0'\not\in \underline{A}(0)$. By (\ref{dia0_05}) and (\ref{dia0_08})  we get
\begin{equation}\label{dia0_09}
x_l(T_1+1)\leq x_{i_0'}(T_1+1)<x_l(0).
\end{equation}
Let $\mathcal{L}_l(s)$ denote the set of the agents in $C_i(x(0))$  whose opinions at time $s$ are less than $x_l(s)$, i.e.,
\begin{eqnarray*}\label{dia0_10}
\mathcal{L}_l(s):=\{m\in C_i(x(0)): x_m(s)<x_l(s)\}.
\end{eqnarray*}
By (\ref{dia0_09}) we have
\begin{eqnarray}\label{dia0_11}
 |\mathcal{L}_l(T_1+1)|\leq |\mathcal{L}_l(0)|-1.
\end{eqnarray}
Let $i_1'$ be the agent in $C_i(x(0))$ which has the smallest opinion within the confidence bound of agent $l$ at time $T_1+1$, i.e.,
\begin{eqnarray*}\label{dia0_12}
&&i_1'=\mathop{\arg\min}_{m\in C_i(x(0))} \{x_m(T_1+1): \\
&&~~~~~~~~~~~~~~~~~~~  |x_l(T_1+1)-x_m(T_1+1)|\leq r_l \}.
\end{eqnarray*}
If $x_{i_1'}(T_1+1)=x_l(T_1+1)$, the agents in $C_i(x(0))$ split into different MC clusters;
otherwise,
let
\begin{multline*}
T_{2}:=T_1+1\\
+\max\left\{\left\lceil\log_{1-\mu} \frac{r_{i_1'}}{x_l(T_1+1)-x_{i_1'}(T_1+1)}\right\rceil,0\right\},
\end{multline*}
and choose $\{i_1',l\}$ as the agent pair for opinion update at times $T_1+1,T_1+2,\ldots,T_2$.

If $i_1'\in \underline{A}(0)$, similar to case I we get $|\underline{A}(T_2+1)|=|\underline{A}(0)|-1$.

If $i_1'\not\in \underline{A}(0)$, similar to (\ref{dia0_11}) we have
\begin{eqnarray}\label{dia0_13}
|\mathcal{L}_l(T_2+1)|\leq |\mathcal{L}_l(T_1+1)|-1.
\end{eqnarray}
Repeat the above process until the agents in $C_i(x(0))$ split into different MC clusters, or $|\underline{A}(T_p+1)|=|\underline{A}(0)|-1$ for some positive integer $p$.
By (\ref{dia0_11})-(\ref{dia0_13}) we get that
\begin{eqnarray*}\label{dia0_14}
p\leq |\mathcal{L}_l(0)|-|\underline{A}(0)|+1\leq |C_i(x(0))|-|\underline{A}(0)|.
\end{eqnarray*}
From this inequality and the definition of $T_1,T_2,\ldots$ we have
\begin{eqnarray}\label{dia0_15}
 T_p+1 \leq \left(|C_i(x(0))|-|\underline{A}(0)|\right)\left(1+\lceil\log_{1-\mu} r_{\min}^i/ r_{\max}^i \rceil \right).\nonumber\\
\end{eqnarray}

Let $t_1$ be the minimal time such that E1 or E2 happens. By (\ref{dia0_08_2}) and (\ref{dia0_15})
we have
\begin{equation}\label{dia0_16}
  t_1\leq \left(|C_i(x(0))|-|\underline{A}(0)|\right)\left(1+\lceil\log_{1-\mu} r_{\min}^i/ r_{\max}^i \rceil\right) .
\end{equation}
If E1 happens at time $t_1$, our result i) holds; otherwise, we need to carry out the following Step 2.

Step 2: For $s\geq t_1$ we control the agent $l$ moves toward the right until E1 or one of the following two events happens:\\
({\bf{E3}}) $x_l(s)\geq x_j(0)+ r_{\min}^i/2$; \\
({\bf{E4}}) $\max_{m\in C_i(x(0))} x_m(s)\leq x_k(0)-(1-\mu)^2 r_{\min}^i$;\\

For $s\geq t_1$, let $i_{s}'$ be the agent in $C_i(x(0))$ which has the biggest opinion within the confidence bound of agent $l$ at time $s$, i.e.,
\begin{eqnarray*}\label{dia0_17}
&&i_{s}'=\mathop{\arg\max}_{m\in C_i(x(0))} \{x_m(s):|x_l(s)-x_m(s)|\leq r_l \}.
\end{eqnarray*}
Choose $\{i_{s}',l\}$ as the agent pair for opinion update, until at least one  of the events E1, E3, and E4 happens.
Let $t_2$ be the minimal time that  E1, E3, or E4 happens.
For $s\in [t_1,t_2)$, since E1 and E4 do not happen at time $s$,
\begin{eqnarray*}\label{dia0_18}
x_l(s+1)=(1-\mu)x_l(s)+\mu x_{i_{s}'}(s)>x_l(s).
\end{eqnarray*}
By the similar method as Step 1, each agent in $ C_i(x(0))\backslash
(\underline{A}(t_1)\cup \{l\})$ can be chosen at most $1+\lceil\log_{1-\mu}
r_{\min}^i/r_{\max}^i\rceil$ times for opinion update during
$[t_1,t_2)$. Then,
\begin{eqnarray}\label{dia0_19}
&&t_2-t_1   \nonumber\\
&& \leq \left(|C_i(x(0))|-|\underline{A}(t_1)|-1\right)\left(1+\lceil\log_{1-\mu}
r_{\min}^i/r_{\max}^i\rceil \right)\nonumber\\
&&  =\left(|C_i(x(0))|-|\underline{A}(0)|\right)\left(1+\lceil \log_{1-\mu}
r_{\min}^i/r_{\max}^i \rceil \right).\nonumber\\
\end{eqnarray}
If E4 happens,  Lemma~\ref{mcc_convex} implies
\begin{eqnarray*}
\begin{aligned}
&~~~  \max_{M,m\in C_i(x(0))} [x_M(t_2)- x_m(t_2)]\\
&~~~  \leq \max_{M\in C_i(x(0))}x_M(t_2)- x_j(0)\\
&~~~  \leq x_k(0)- x_j(0)-(1-\mu)^2 r_{\min}^i,
\end{aligned}
\end{eqnarray*}
which indicates our result ii) holds;
if E1 happens, our result i) holds at time $t_2$;
otherwise, we need to carry out next Step.\\
... ...\\[0.1in]
Step $2m+1$: For $s\geq t_{2m}$, we use the similar control method as Step
1. Let $t_{2m+1}$ be the minimal time such that E1 happens or
$|\underline{A}(t_{2m+1})|=|\underline{A}(t_{2m-1})|-1$. Similar to
(\ref{dia0_16}) we have
 \begin{eqnarray}\label{dia0_21}
&&t_{2m+1}-t_{2m}\\
&&\leq  \left(|C_i(x(0))|-|\underline{A}(t_{2m-1})|\right)\left(1+\lceil \log_{1-\mu}
r_{\min}^i/r_{\max}^i\rceil \right)\nonumber\\
&&= \left(|C_i(x(0))|-|\underline{A}(0)|+m\right)\left(1+\lceil \log_{1-\mu}
r_{\min}^i/r_{\max}^i\rceil \right).\nonumber
\end{eqnarray}

Step $2m+2$: For $s\geq t_{2m+1}$,
we use the similar control method as Step 2. Let $t_{2m+2}$ be the minimal time such that E1, E3, or E4 happens. Similar to (\ref{dia0_19}) we have
 \begin{eqnarray}\label{dia0_22}
&&t_{2m+2}-t_{2m+1}\\
&&\leq  \left(|C_i(x(0))|-|\underline{A}(t_{2m+1})|-1\right)\left(1+\lceil \log_{1-\mu}
r_{\min}^i/r_{\max}^i\rceil \right)\nonumber\\
&&=\left(|C_i(x(0))|-|\underline{A}(0)|+m\right)\left(1+\lceil \log_{1-\mu}
r_{\min}^i/r_{\max}^i\rceil \right).\nonumber
\end{eqnarray}

The above process will end at Step $2|\underline{A}(0)|-1$ because $\underline{A}(t_{2|\underline{A}(0)|-1})=\emptyset$.  By
 Lemma \ref{mcc_convex} and the definition of $\underline{A}(s)$ we have
 \begin{eqnarray}\label{dia0_23}
\begin{aligned}
&\max_{M,m\in C_i(x(0))} [x_M(t_{2|\underline{A}(0)|-1})- x_m(t_{2|\underline{A}(0)|-1})]\\
&\leq x_k(0)- \min_{m\in C_i(x(0))}x_m(t_{2|\underline{A}(0)|-1})\\
&\leq x_k(0)- x_j(0)-(1-\mu)^2 r_{\min}^i,
\end{aligned}
\end{eqnarray}
 which indicates our result ii) holds when $t^*=t_{2|\underline{A}(0)|-1}$. Set $t_0:=0$. By (\ref{dia0_22}) and (\ref{dia0_23}) we have
 \begin{eqnarray*}\label{dia0_24}
\begin{aligned}
&t_{2|\underline{A}(0)|-1}\\
&=\sum_{m=0}^{|\underline{A}(0)|-2}\left(t_{2m+2}-t_{2m}\right)+t_{2|\underline{A}(0)|-1}-t_{2|\underline{A}(0)|-2}\\
&\leq \Big(\sum_{m=0}^{|\underline{A}(0)|-2}2\left(|C_i(x(0))|-|\underline{A}(0)|+m\right)\\
&~~~~+|C_i(x(0))|-1\Big)\left(1+\lceil \log_{1-\mu}
r_{\min}^i/r_{\max}^i\rceil \right)\\
&=\big[(2|\underline{A}(0)|-1)|C_i(x(0))|+(-|\underline{A}(0)|-1)|\underline{A}(0)|+1 \big]\\
&~~~~\times \left(1+\lceil \log_{1-\mu}
r_{\min}^i/r_{\max}^i\rceil\right)\\
&\leq (|C_i(x(0))|-1)^2 \left(1+\lceil \log_{1-\mu}
r_{\min}^i/r_{\max}^i\rceil \right),
\end{aligned}
\end{eqnarray*}
where the last inequality uses the fact that $|\underline{A}(0)|\leq |C_i(x(0))|-1$.

For the case when  $x_l(0)< [x_k(0)+x_j(0)]/2$, we can set
$$\bar{A}(s):=\{m\in C_i(x(0)): x_m(s)> x_k(0)-(1-\mu)^2 r_{\min}^i\},$$
and use the similar method as the case $x_l(0)\geq [x_k(0)+x_j(0)]/2$ to control $\bar{A}(s)$ becomes empty.
\oprocend

\end{document}